\documentstyle[12pt,epsf]{article}

\newcommand{\be}{\begin{equation}}
\newcommand{\ee}{\end{equation}}
\newcommand{\ba}{\begin{array}{c}}
\newcommand{\ea}{\end{array}}
\newcommand{\bqa}{\begin{eqnarray}}
\newcommand{\eqa}{\end{eqnarray}}
\begin{document}

\begin{flushright}
\end{flushright}
\vspace*{2.0cm}

\begin{center}
{\Large\bf A Numerical Analysis to the { $\pi$} and { $K$}
Coupled--Channel Scalar Form-factor}
\\[10mm]

{\sc Wei Liu, Hanqing Zheng and Xiao-Lin Chen}
\\[5mm]
{\it 1) Department of Physics, Peking University, Beijing 100871,
P.~R.~China}\\

\end{center}

\begin{abstract}
A  numerical analysis to the scalar form-factor in the $\pi\pi$
and $KK$ coupled--channel system is  made by  solving the
coupled-channel dispersive integral equations, using the iteration
method. The solutions are found not unique. Physical application to
the $\pi\pi$ central production in the $pp\to pp\pi\pi$ process is
discussed based upon the numerical solutions we found.
\end{abstract}


\vspace{1cm}

The I=J=0 channel $\pi\pi $ interactions are of great physical
interests. Because the interaction of the I=J=0 channel is very
strong, the input bare singularities for a given model can be
severely renormalized and distorted by the strong attractive force
and extra dynamical singularities may be generated~\cite{LMZ2}.
Therefore the I=J=0 channel affords an ideal test ground for
models of strong interactions. Also, the lightest glueball is
expected to lie in the I=J=0 channel, as predicted by the lattice
QCD calculations. Therefore a detailed study to the dynamics in
this channel becomes especially interesting. The low energy I=J=0
$\pi\pi$ system manifests itself in various production processes
intensively measured by experiments, $i.e.,$ from $\pi N\to \pi\pi
N$, $\gamma\gamma \to \pi\pi$ to $J/\Psi\to \phi \pi\pi$,  etc.,
with the center of mass energy, $\sqrt{s}$, ranging from the
$\pi\pi$ threshold to a few GeV. When $\sqrt{s}$ exceeds the
$K\bar K$ threshold, a single--channel analysis to the $\pi\pi$
system becomes inadequate, and instead, a coupled--channel
analysis of $\pi\pi$ and $K\bar K$ system has to be made. Among
various $\pi\pi$ production processes particularly interesting
cases are those I=J=0 final states which are generated weakly. In
such circumstances, the final state particles will not scatter
back  to the initial states, and therefore the complicated
dynamics involved is considerably simplified without loss of
information on the I=J=0 final state interactions. For example, in
such a case the Watson--Migdal's theorem on final state
interactions applies. The $\pi\pi$ final states ``weakly''
produced can be again categorized into two classes: One is that
the $\pi\pi$ production vertex contains left--hand singularities
like in the case $\gamma\gamma\to \pi\pi$~\footnote{A recent
publication on related subject can be found in Ref.~\cite{oller}.}
and the another is not, like in the case $K\to\pi\pi$ and in the
$\pi\pi$ central production process $pp\to pp\pi\pi$. Physical
situation in the absence of left hand singularities is further
simplified, since the dynamical complexity from the production
vertex is removed which would otherwise disturb our analysis on
the $\pi\pi$ (final state) interaction itself. In such a
simplified situation, it then becomes  reasonable to assume that
the $\pi\pi$ production amplitude is factorized as a product of a
form-factor--like quantity, which we denote as $A$, containing all
the dynamical singularities from the right--hand cut from $\pi\pi$
final state interactions, and the production vertex  which is a
smooth analytic function of $s$ on the complex $s$--plane except
possibly at infinity since it contains neither the left--hand nor
the right--hand singularities.

For a coupled-channel system of $\pi\pi$ and $KK$, the spectral
representation of the form-factor, ${\bf A}\equiv (A_1, A_2)$,
satisfies the following relation,
\begin{equation}
\begin{array}{lcr}
  {\mathrm Im} A_1 & = & A_1\rho_1T_{11}^++A_2\rho_2T_{21}^+\ ,    \\
  {\mathrm Im} A_2 & = & A_2\rho_2T_{22}^++A_1\rho_1T_{12}^+\ ,
\end{array} \label{uni1}
\end{equation}
 whereas the unitarity relation of the scattering matrix ${\bf T}$ reads,
\be
  {\rm Im} \mathbf{T} = \mathbf{T} \rho \mathbf{T}^+\ ,
  \label{uni2}
\ee where ${\bf\rho}\equiv diag(\rho_1, \rho_2)$ is the matrix of
the kinetic phase-space factor. The $\mathbf{T}$ matrix may
contain left--hand cut but $\mathbf{A}$ does not. Especially,
$\mathbf{A}$ is analytic on the entire physical sheet of the
complex $s$ plane except on the cut along the real positive axis
starting from $2\pi$ threshold. The form-factor $\mathbf{A}$ has
the same analytic structure as the scalar
form-factor~\footnote{For the definition of the scalar
form-factor, see for example Ref.~\cite{oller}.} and the two are
different only up to a polynomial. Eqs.~(\ref{uni1}) and
(\ref{uni2}) are assumed to be correct down to the lowest
threshold, in the absence of anomalous thresholds. That means when
$4m_\pi^2\le s\le 4m_K^2$ Eq.~(\ref{uni1}) takes the form,
\begin{equation}
\begin{array}{lcr}
  {\mathrm Im} A_1 & = & A_1\rho_1T_{11}^+\ ,    \\
  {\mathrm Im} A_2 & = & A_1\rho_1T_{12}^+\ .
\end{array} \label{uni1b}
\end{equation}

From previous discussions, it is realized that the physical
problem of studying final state interactions in the production
process without left--hand singularities is reduced to the
mathematical problem of solving Eq.~(\ref{uni1}) (and
Eq.~(\ref{uni1b})), provided that the T matrix is known. In the
single channel case the analytic solution of the form factor can
be obtained. The spectral representation of the form-factor is
given by the first equation in Eq.~(\ref{uni1b}) from which the
classical Omn\`es solution can be established,
\begin{eqnarray}
  A(s) & = & P(s) \exp \left( \frac{s}{\pi}
             \int_{4m_{\pi}^2}^{\infty} \frac{\delta_\pi(s')}
             {s'(s'-s-i\epsilon)}ds' \right)\ , \label{FOM}
\end{eqnarray}
where $P(s)$ is a polynomial and $\delta_\pi$ is the $\pi\pi$
scattering phase. The Omn\`es solution is remarkable in relating
the form-factor to  $\delta_\pi$. The Eq.~(\ref{FOM}) has been
used  in Ref.~\cite{LMZ} to study the  final state interactions in
$K\to 2\pi$ and $pp\to pp\pi\pi$ systems. In the coupled--channel
case, however, no analytic solutions of Eq.~(\ref{uni1}) can be
obtained in general.
 In the following
we will instead study the coupled--channel system by numerical
method~\cite{dgl}.

Since the coupled--channel form-factor $\mathbf{A}$ is analytic on
the physical sheet of the complex $s$ plane except for the cut
along the real axis,  assuming the $\mathbf{T}$ matrix is known we
can search for solutions of the amplitude $\mathbf{A}$ in
Eq.~(\ref{uni1}) by solving the following dispersion relation:
\be\label{IE} {\bf A}={1\over \pi}\int_R{ {\bf A\rho T^+}(s')\over
s'-s-i\epsilon}ds'\ , \ee where the integration is performed on
the unitarity cut $R$, starting from $4m_\pi^2$ to $\infty$. The
Eq. (\ref{IE}), according to Muskhelishvili~\cite{muskhelishvili},
contains two fundamental solutions, ${\bf \phi_1}$ and ${\bf
\phi_2}$. Assuming that $\phi_n$ behaves as $\phi_n\to
s^{-\chi_n}$ as $s\to \infty$ one has
\be
  \sum_n\chi_n = {1\over 2\pi}[arg\, det{\bf S}]_R=
  {1\over \pi}(\delta_\pi(\infty)+\delta_K(\infty))\ ,
\ee where  ${\bf S}$ is the coupled--channel  $S$ matrix.
According to \cite{muskhelishvili}, any solution of the integral
equation (\ref{IE}) can be written as a linear composition of the
two fundamental solutions,
\be
  \phi=\sum_n P_n(s)\phi_n\ \quad , \quad n=1,2
\ee where $P_n(s)$ are polynomials of $s$. The polynomials are not
determined  from analyticity alone. Other physical input has to be
implemented to fix their coefficients. For example, chiral
perturbation theory can afford an  expansion of the scalar
form-factor in powers of $s$ when $s$ is small~\cite{GM}. Since
there can be many solutions of Eq.~(\ref{IE}) and we notice that,
since the numerical integration in Eq.~(\ref{IE}) has to be
truncated somewhere (denoted as $\Lambda$ below), all the
information on the asymptotic behaviour are lost. As a
consequence,  it is difficult to distinguish  the so--called
fundamental solutions from  others, since the difference between
the two essentially comes from their asymptotic behaviours.
Therefore in the present numerical scheme instead of searching for
the fundamental solutions,
 we follow the recipe of
Ref.~\cite{dgl}, that is to search for two linearly independent
solutions, ${\bf A^{1}}$ and ${\bf A^{2}}$ which are
normalized at $s=0$ as,
\be\label{b1}
  A_1^{1}(0)=1\ , \quad A_2^{1}(0)=0\ ,
\ee and \be\label{b2}
  A_1^{2}(0)=0\ , \quad A_2^{2}(0)=1\ .
\ee

\begin{figure}[htb]
\begin{center}
\mbox{\epsfysize=60mm \epsffile{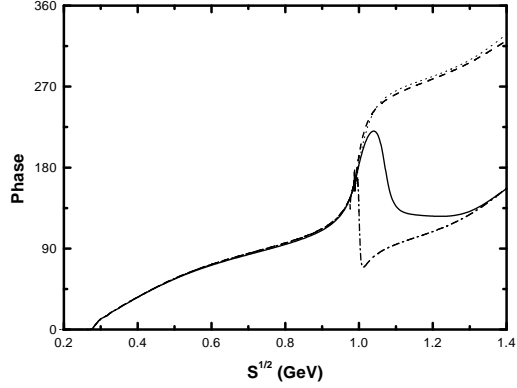}} \hspace*{15mm}
\caption{\label{fig1}   The solid, dashed and dot--dashed lines
represent the  phases of different solutions of the $A_1^{1}$
amplitudes, the corresponding magnitude of the amplitudes are
depicted in Fig.~\ref{fig2}. The dotted  line represents the phase
of $T_{11}$ from the K3 fit of Ref.~\protect\cite{AMP}.}
\end{center}
\end{figure}

\begin{figure}[htb]
\begin{center}
\mbox{\epsfysize=60mm \epsffile{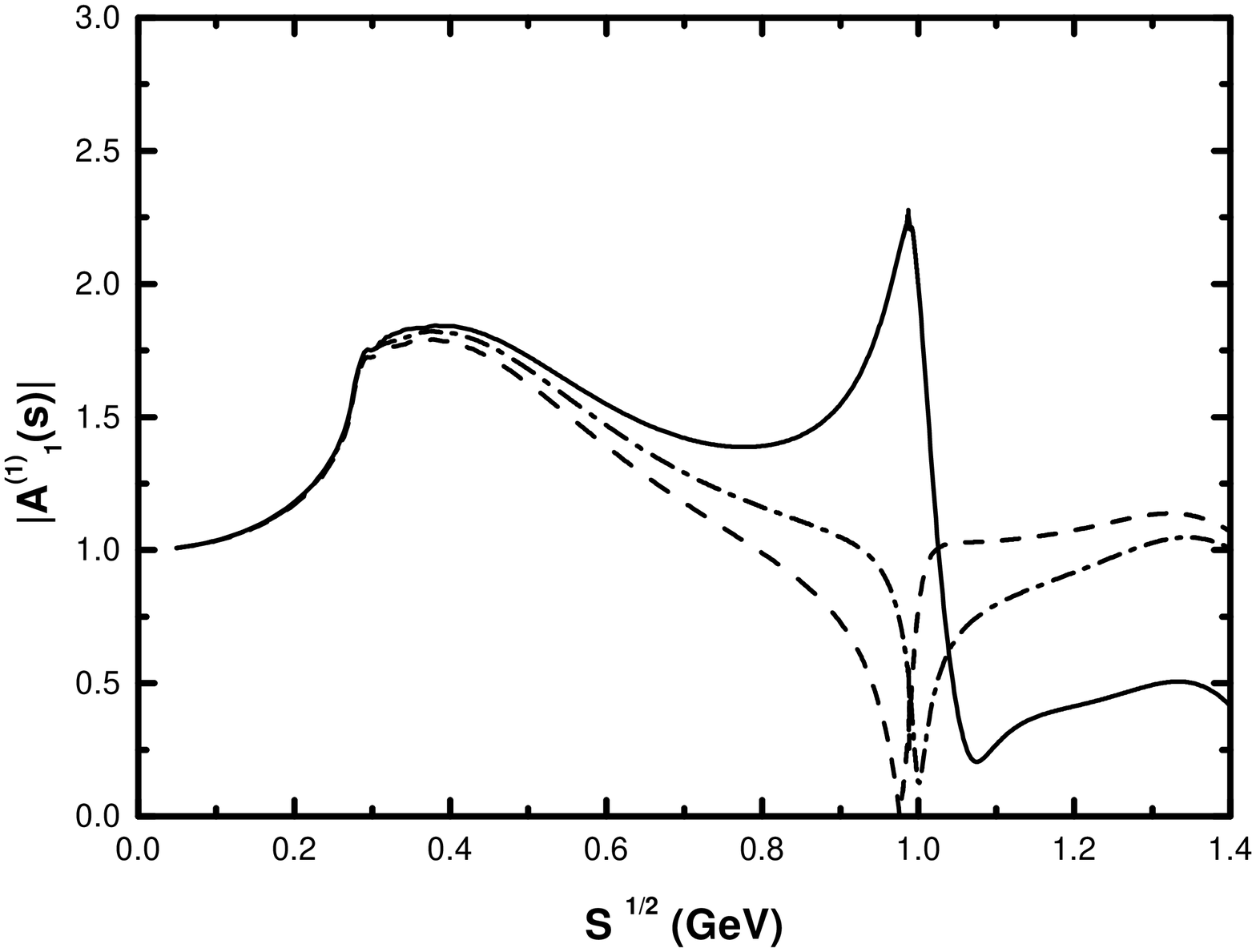}} \hspace*{15mm}
\caption{\label{fig2}   The magnitude of the different solutions
of the $A_1^{1}$ amplitude.}
\end{center}
\end{figure}

\begin{figure}[htb]
\begin{center}
\mbox{\epsfxsize=100mm \epsffile{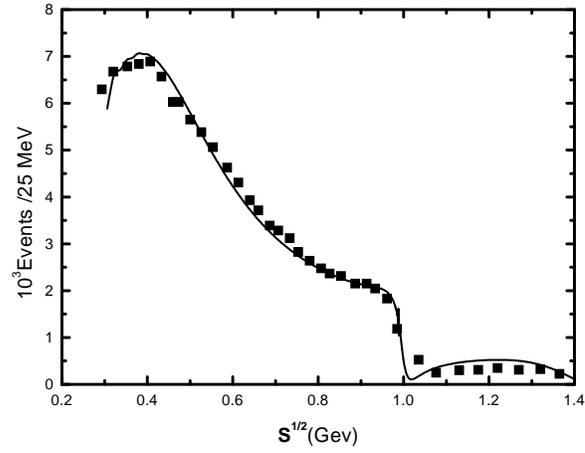}} \hspace*{15mm}
 \caption{\label{fig3} The
effective mass distribution of  pion pairs in $pp\to pp\pi\pi$ vs.
$M=\protect\sqrt{s}$. The data are from \protect\cite{AFS}.}
\end{center}
\end{figure}

In the following we use the $\pi\pi$ and $KK$ coupled--channel fit
of the T matrix from Au, Morgen and Penington~\cite{AMP} as an
educative  example to solve the coupled--channel dispersive
integral equation (\ref{IE}). The integral in Eq.~(\ref{IE}) is
truncated at $\Lambda\simeq 1.5GeV$. We find that the influence of
the cutoff is rather local, i.e., it has little effects on the
behaviour of the solution in a large region of $s$, even close to
$\Lambda$. One can also use a mild regulation function instead of
the truncation at $\Lambda$ and the result is essentially the same
except at $s\simeq\Lambda$. We use the iteration method to solve
Eq.~(\ref{IE}). That is, \be\label{IEI} {\bf A^{(n+1)}}(s)={\bf
A}(0)+{s\over \pi}\int^\Lambda_{4m_\pi^2}{ {\rm Real}\left[{\bf
A^{(n)}(s')\rho T^+}(s')\right]\over s'(s'-s-i\epsilon)}ds'\ , \ee
where \be\label{a} {\bf A}(0)=(1,0) \,\,\, {\rm or}\,\,\, (0,1)\ .
\ee
 A once--subtracted
form of the dispersive integral in the above Eq.~(\ref{IEI}) is
helpful in incorporating the boundary conditions, Eqs.~(\ref{b1})
and (\ref{b2}). The routine converges rather rapidly (after about
20 steps), which confirms the claim in Ref.~\cite{dgl} and the
solutions are not unique due to the reason already mentioned
above. This means that the iteration may converge to different
solutions depending on different initial values for the iteration.
In Fig.~\ref{fig1} and Fig.~\ref{fig2} a few examples generated
from our numerical recipe are shown. We see from Fig.~\ref{fig1}
that all the phases are identical to the phase of $T_{11}$ below
the $K\bar K $ threshold as required by the final state theorem.
Above the $K\bar K $  threshold, however, the phases from
different solutions  of $A_1$ can be very different and also
deviate from the phase of $T_{11}$. In Fig.~\ref{fig2} the
corresponding magnitude of the different solutions is also shown.
We see that around $1GeV$ region there can be zeros or dips to
compensate the peak generated by $f_0(980)$. The physical
discussion on the necessity to introduce the protective zero
appearing in the $\pi\pi$ production processes can be found in
Ref.~\cite{LMZ,zero}. In the present numerical approach there is
generally no difficulty to pick suitable solutions, from all, to
fit experimental data. An example is shown in Fig.~\ref{fig3}
where the $\pi\pi$ production cross-section in the $pp\to
pp\pi\pi$ process is reproduced.
 The cross-section  can be
expressed by \cite{MP84},
\begin{eqnarray}
  \frac{d\sigma}{d \sqrt{s}} \sim \frac{(s-4 m_{\pi}^2)^{1/2}}{s^{3/2}}
  |F(s)|^2\ ,
\end{eqnarray}
where $F(s)$ is a linear combination of solutions $A_1^{1}$ and
$A_1^{2}$, \be\label{F} F(s)=\alpha_1 A_1^{1}+\alpha_2 A_1^{2}\ .
\ee It is assumed here that $\pi\pi$ produced  in the $pp\to
pp\pi\pi$ process are generated from the fusion of Pomeron pairs
and no left--hand singularity is involved in the production
vertex. Therefore $\alpha_1$ and $\alpha_2$ (which contain
information about the $\pi\pi$ production vertex) appearing in the
above equation are expected to be polynomials with a weak
dependence on $s$.

In fact, the T matrix given in Ref.~\cite{AMP} is obtained by a
multi-pole K matrix fit in which the background effects including
the left--hand cut one are simulated by a polynomial. Therefore,
the T matrix does not contain any left--hand singularity. Under
such an approximation it is not really necessary to search for the
numerical solutions of $A_1$ in order to make use of
Eq.~(\ref{F}). Since $A_1$ has to be a linear combination of
$T_{11}$ and $T_{21}$ when neglecting the left--hand singularities
in $T$, one can use instead of Eq.~(\ref{F}) the following form,
\be\label{F'} F(s)=\alpha'_1 T_{11}+\alpha'_2 T_{21}\ , \ee to fit
the experimental data, as originally done in Ref.~\cite{AMP}. The
method presented in the current note applies to more general cases
when the fine details of the left--hand singularities of the
scattering $T$ matrix are carefully taken into account. Also our
program can be easily extended to the situation when the
left--hand cut of the production vertex is included. We will
investigate these more realistic cases in future.

\noindent {$\bf Acknowledgement$:} One of the authors, H.Z. would
like to thank M.~Locher and V.~Markushin  for helpful discussions.
He is supported in part by National Natural Science Foundation of
China under grant No.~19775005.



\end{document}